\documentclass[preprint,aps,nofootinbib,showpacs,amsfonts,epsf]{revtex4}

\input{epsf.tex}

\newcommand{\E}{{\cal{E}}}
\newcommand{\s}{\sigma}

\renewcommand{\a}{\alpha}

\newcommand{\be}{\begin{equation}}
\newcommand{\ee}{\end{equation}}
\newcommand{\bea}{\begin{eqnarray}}
\newcommand{\eea}{\end{eqnarray}}
\newcommand{\ba}{\begin{array}}
\newcommand{\ea}{\end{array}}
\def\J#1#2#3#4{{#1} {\bf #2}, #3 (#4)}
\def\PRD{Phys. Rev. D}
\def\PR{Phys. Rev.}
\def\PRL{Phys. Rev. Lett.}
\def\RMP{Rev. Mod. Phys.}

\def\AA{Astron. Astrophys.}

\def\MNRAS{Mon. Not. R. Astron. Soc.}
\def\JMP{J. Math. Phys.}

\def\CQG{Class. Quantum Grav.}

\begin{document}
\draft
\title{Simple metric for a magnetized, spinning, deformed mass}

\author{V.~S.~Manko$^\dagger$ and E.~Ruiz$^\ddagger$ }
\address{$^\dagger$Departamento de F\'\i sica, Centro de Investigaci\'on y
de Estudios Avanzados del IPN, A.P. 14-740, 07000 Ciudad de
M\'exico, Mexico\\$^\ddagger$Instituto Universitario de F\'{i}sica
Fundamental y Matem\'aticas, Universidad de Salamanca, 37008
Salamanca, Spain}

\begin{abstract}
We present and discuss a 4-parameter stationary axisymmetric
solution of the Einstein-Maxwell equations able to describe the
exterior field of a rotating magnetized deformed mass. The
solution arises as a system of two overlapping corotating
magnetized non-equal black holes or hyperextreme disks and we
write it in a concise explicit form very suitable for concrete
astrophysical applications. An interesting peculiar feature of
this solution is that its first four electric multipole moments
are zeros; it also has a non-trivial extreme limit which we
elaborate completely in terms of four polynomial factors. We
speculate that the formation of the binary configurations of this
type, which is accompanied by a drastic change of the system's
total angular momentum due to strong dragging effects, might be
one of the mechanisms giving birth to relativistic jets in the
galactic nuclei.
\end{abstract}


\maketitle

\section{Introduction}

During the last decade there has been considerable interest in the
observational confirmation of the nature of the known black hole
(BH) candidates as yet another possible test of general relativity
\cite{Bam,BBC}. It is clear that for being able to recognize a
Kerr BH \cite{Ker} by analyzing the data obtained in a real
astronomical observation, it is also necessary to know well the
properties of the non-Kerr spacetimes which could only slightly
differ from those of a BH. Moreover, proposals for the study of
the properties of black holes with electromagnetic radiation
\cite{Joh,Bam2} make desirable the knowledge of simple models for
compact magnetized objects permitting a clear physical
interpretation. Such models may arise in particular from the
extended 2-soliton electrovac solution with equatorial symmetry
\cite{MMR} determining a large 6-parameter family of the two-body
configurations and permitting to analytically approximate the
exterior fields of compact astrophysical objects. Note that in
order to form a configuration with reflection symmetry, a two-body
system must be composed either of two separated identical
corotating constituents, or by two nonequal overlapping
constituents with a common center and with a corotation or a
counterrotation. While the former type of two-body systems is more
habitual for the analysis and studies, the systems of the latter
type as a rule are given no attention at all because their
interpretation may look clumsy from the black hole point of view.
At the same time, the use of overlapping black holes for modeling
compact objects of astrophysical interest seems to us very
natural, and for instance a pair of superposed Schwarzschild black
holes with a common center could be a good model of a static
deformed mass. Similarly, the double-Reissner-Nordstr\"om solution
\cite{Man} in which the separation distance $R$ is set equal to
zero would describe a simple model of a charged static deformed
mass; besides, as can be easily checked, in the $R=0$ limit this
solution simplifies considerably and becomes a static
specialization of the solution \cite{MMR}. In the present paper we
will explore further the second type of the equatorially symmetric
configurations and consider a 4-parameter electrovac metric for a
magnetized rotating deformed mass which also could be regarded as
representing two overlapping black-hole constituents and which was
identified with the help of a recent paper on the charged rotating
masses \cite{Cab}. The solution will be shown to have a simple
form, with various distinctive features and interesting limits,
which makes it potentially suitable for the use in astrophysical
applications.

Our paper is organized as follows. In the next section we present
the 4-parameter electrovac solution, the corresponding complete
metric and expression of the magnetic potential. Here we also
analyze the sub- and hyperextreme cases of the solution and its
multipole structure. In Sec.~III we consider interesting
specializations of the solution and work out the solution's
extreme limit. Concluding remarks are given in Sec.~IV.

\section{The 4-parameter electrovac solution and metric functions}

Let us begin this section with reminding that the general
equatorially symmetric 2-soliton electrovac solution \cite{MMR}
(henceforth referred to as the MMR solution), obtained with the
aid of Sibgatullin's integral method \cite{Sib,MSi}, is defined by
the axis data
\be e(z)=\frac{(z-m-ia)(z+ib)+k} {(z+m-ia)(z+ib)+k}, \quad
f(z)=\frac{qz+ic} {(z+m-ia)(z+ib)+k}, \label{axisMMR} \ee
where $e(z)$ and $f(z)$ are the expressions of the Ernst complex
potentials $\E$ and $\Phi$ \cite{Ern} on the upper part of the
symmetry $z$-axis; the six arbitrary real parameters entering
(\ref{axisMMR}) are $m$, $a$, $b$, $k$, $q$ and $c$.

The 4-parameter specialization of the MMR solution which we are
going to report in this paper is determined by the following
simple choice of the parameters in (\ref{axisMMR}):
\be b=a, \quad k=m_1m_2-\mu^2, \quad q=0, \quad c=m\mu,
\label{choice} \ee
the mass parameters $m_1$ and $m_2$ being such that $m_1+m_2=m$;
the charge parameter $q$ is set equal to zero because of its
irrelevance for astrophysical applications, and hence the
electromagnetic field in this solution is defined solely by the
magnetic dipole parameter $\mu$. Then the axis data determining
this particular case take the form
\bea e(z)=\frac{(z-m_1)(z-m_2)-ia(m_1+m_2)+a^2-\mu^2}
{(z+m_1)(z+m_2)+ia(m_1+m_2)+a^2-\mu^2}, \nonumber\\
f(z)=\frac{i\mu(m_1+m_2)} {(z+m_1)(z+m_2)+ia(m_1+m_2)+a^2-\mu^2},
\label{axis4p} \eea
where the four arbitrary real parameters are $m_1$, $m_2$, $a$ and
$\mu$. The particular parameter choice (\ref{choice}) occurred to
us when we noticed that the metric for two unequal counterrotating
charged masses \cite{Cab} becomes, in the limit $R=0$, a member of
the MMR solution.

An attractive feature of the data (\ref{axis4p}) is that the
algebraic equation
\be e(z)+\bar e(z)+2f(z)\bar f(z)=0 \label{aleq}\ee
(the bars over symbols denote complex conjugation), which plays an
important role in Sibgatullin's method, yields for this case four
very simple roots $\a_n$, namely,
\be \a_1=-\a_2=\sqrt{m_1^2-a^2+\mu^2}\equiv\s_1, \quad
\a_3=-\a_4=\sqrt{m_2^2-a^2+\mu^2}\equiv\s_2, \label{alfas} \ee
which in turn define the following four functions of the
Weyl-Papapetrou coordinates ($\rho,z$):
\be R_\pm=\sqrt{\rho^2+(z\pm\s_1)^2},  \quad
r_\pm=\sqrt{\rho^2+(z\pm\s_2)^2}. \label{Rr} \ee

Formulas (\ref{axis4p}), (\ref{alfas}) and (\ref{Rr}) permit one
to construct, by purely algebraic manipulations, the Ernst
potentials $\E$ and $\Phi$, as well as the corresponding metric
functions $f$, $\gamma$ and $\omega$ of the line element
\be d s^2=f^{-1}[e^{2\gamma}(d\rho^2+d z^2)+\rho^2 d\varphi^2]-f(d
t-\omega d\varphi)^2, \label{Papa} \ee
using only the determinantal formulas of the paper \cite{RMM}. In
our concrete case the desired expressions can be also worked out
from the respective formulas of the MMR
solution.\footnote{Unfortunately, the general formulas of the
paper \cite{Cab} are not helpful for elaborating the $R=0$ limit
because of the misprints crept into that paper, so that
Eqs.~(\ref{EF}) and (\ref{mf}) have been worked out with the aid
of our computer codes developed for the MMR solution.} As a
result, the potentials $\E$ and $\Phi$ of the 4-parameter solution
can be shown to have the form
\bea \E&=&\frac{A-B}{A+B}, \quad \Phi=\frac{C}{A+B}, \nonumber\\
A&=&\s_1\s_2[(m_1^2+m_2^2)(R_++R_-)(r_++r_-)
-4m_1m_2(R_+R_-+r_+r_-)] \nonumber\\ && -(m_2^2\s_1^2+m_1^2\s_2^2)
(R_+-R_-)(r_+-r_-) +ia(m_1^2-m_2^2)
\nonumber\\
&& \times[\s_1(R_++R_-)(r_+-r_-)-\s_2(R_+-R_-)(r_++r_-)],
\nonumber\\
B&=&-2(m_1^2-m_2^2)\{\s_1\s_2[m_2(R_++R_-)-m_1(r_++r_-)]
+ia[m_2\s_2(R_+-R_-) \nonumber\\ &&-m_1\s_1(r_+-r_-)]\}, \nonumber\\
C&=&2i\mu(m_1^2-m_2^2) [m_1\s_2(R_+-R_-)-m_2\s_1(r_+-r_-)],
\label{EF} \eea
while for the metrical fields $f$, $\gamma$ and $\omega$ one gets
the following expressions:
\bea f&=&\frac{A\bar A-B\bar B+C\bar C}{(A+B)(\bar A+\bar B)},
\quad e^{2\gamma}=\frac{A\bar A-B\bar B+C\bar C}{K_0\bar
K_0R_+R_-r_+r_-}, \quad \omega=-\frac{{\rm Im}[G(\bar
A+\bar B)+C\bar I]}{A\bar A-B\bar B+C\bar C}, \nonumber\\
G&=&-2zB+(m_1^2-m_2^2)\{\s_1(2m_2^2
+\mu^2)(R_++R_-)(r_+-r_-) \nonumber\\
&&-\s_2(2m_1^2+\mu^2) (R_+-R_-)(r_++r_-)
-2ia(m_1^2-m_2^2)(R_+-R_-)(r_+-r_-) \nonumber\\
&&-2\s_2[2m_2\s_1^2-\mu^2(m_1+m_2)](R_+-R_-) +2\s_1[2m_1\s_2^2
-\mu^2(m_1+m_2)](r_+-r_-) \nonumber\\
&& -4ia\s_1\s_2[m_2(R_++R_-)-m_1(r_++r_-)]\}, \nonumber\\
I&=&-i\mu(m_1+m_2)\Big( \s_1\s_2[(m_1+m_2)(R_++R_-)(r_++r_-)
-4m_1R_+R_--4m_2r_+r_-] \nonumber\\
&&-(m_2\s_1^2+m_1\s_2^2)(R_+-R_-)(r_+-r_-)
+ia(m_1-m_2)[\s_1(R_++R_-)(r_+-r_-) \nonumber\\
&&-\s_2(R_+-R_-)(r_++r_-)]-2(m_1-m_2)
\{\s_1\s_2[(2m_1+m_2)(R_++R_-) \nonumber\\
&&-(m_1+2m_2)(r_++r_-)+2m_1^2-2m_2^2] +ia(m_1+m_2)[\s_2(R_+-R_-)
-\s_1(r_+-r_-)]\}\Big), \nonumber\\ K_0&=&4\s_1\s_2(m_1-m_2)^2.
\label{mf} \eea

It should be also noted that the electric $A_t$ and magnetic
$A_\varphi$ components of the electromagnetic 4-potential defined
by the solution (\ref{EF}) have the form
\be A_t=-{\rm Re}\left(\frac{C}{A+B}\right), \quad A_\varphi={\rm
Im}\left(\frac{I-zC}{A+B}\right), \label{At} \ee
and these formulas complement the description of our particular
4-parameter electrovac spacetime.

Turning now to the discussion of the properties of the solution
(\ref{EF}), we first mention that the form of $\s_1$ and $\s_2$
defined in (\ref{alfas}) clearly shows which type of constituents
may form the two-body configurations described by this solution.
In the subextreme case, the quantities $\s_1$ and $\s_2$ are
real-valued, which means that both $m_1$ and $m_2$ must fulfil the
inequalities $m_1^2>a^2-\mu^2$ and $m_2^2>a^2-\mu^2$. Similarly,
in the hyperextreme case both $\s_1$ and $\s_2$ take pure
imaginary values, which means that $m_1^2<a^2-\mu^2$ and
$m_2^2<a^2-\mu^2$. Since $\s_1\ne\s_2$ generically, we can
suppose, say, that $m_1>m_2$; then the mixed
subextreme-hyperextreme case arises when $m_2^2<a^2-\mu^2<m_1^2$.
These three basic types of the two-body configurations described
by the solution (\ref{EF}) are depicted in Fig.~1. The fact that
the constituents are overlapping can be most easily seen by
setting the rotational parameter $a$ equal to zero in the above
formulas and observing that in this case the solution reduces to
the $R=0$ limit of the asymmetric dihole spacetime considered in
\cite{MRS}.

The calculation of the first five Beig-Simon relativistic
multipole moments \cite{Sim} with the aid of the
Hoenselaers-Perj\'es procedure \cite{HPe} rectified by Sotiriou
and Apostolatos \cite{SAp} yields for the solution (\ref{EF}) the
expressions
\bea P_0&=&m_1+m_2, \quad P_1=ia(m_1+m_2), \quad
P_2=-(m_1+m_2)(m_1m_2+a^2-\mu^2), \nonumber\\
P_3&=&-ia(m_1+m_2)(m_1m_2+a^2-\mu^2), \nonumber\\
P_4&=&(m_1+m_2)(m_1m_2+a^2-\mu^2)^2+\frac{1}{70}
(m_1+m_2)^3(10m_1m_2-7\mu^2), \nonumber\\ Q_0&=&0, \quad
Q_1=i\mu(m_1+m_2), \quad Q_2=0, \quad
Q_3=-i\mu(m_1+m_2)(m_1m_2+a^2-\mu^2), \nonumber\\
Q_4&=&-\frac{1}{10}a\mu(m_1+m_2)^3, \label{mult} \eea
whence it follows that the parameters $m_1$ and $m_2$ can be
associated with the individual masses of the first and second
constituent, respectively. Moreover, the expression of the total
angular momentum $P_1$ indicates that the constituents are
corotating with the same angular momentum per unit mass ratio:
$j_1/m_1=j_2/m_2=a$, $j_1$ and $j_2$ being angular momenta of the
first and second constituent, respectively. A surprising feature
of the electromagnetic moments $Q_n$ is that the first four
electric multipoles (these are represented by the real parts of
the respective $Q_n$) are zeros, the first nonzero electric moment
being the hexadecapole one.

\section{Physically interesting limits of the solution}

The 4-parameter solution (\ref{EF}) has various physically
interesting limits which we will briefly consider below.

\subsection{The static limit}

When $a=0$, the solution describes a static deformed mass endowed
with magnetic dipole moment, and it coincides, as was already
mentioned earlier, with the $R=0$ specialization of the asymmetric
dihole solution considered in \cite{MRS}. In this case
$\s_1=\sqrt{m_1^2+\mu^2}$, $\s_2=\sqrt{m_2^2+\mu^2}$, which means
that only the subextreme type of overlapping constituents is
possible (see Fig.~1(a)). An interesting particular case of this
magnetostatic solution is the magnetized Schwarzschild metric
which arises by further setting to zero one of the parameters
$m_1$ or $m_2$, and its explicit form is the following
($|C|^2\equiv C\bar C$):
\bea \E&=&\frac{A-B}{A+B}, \quad \Phi=\frac{C}{A+B}, \quad
A_\varphi=-\mu+\frac{2\mu I}{A+B}, \nonumber\\
f&=&\frac{A^2-B^2+|C|^2}{(A+B)^2}, \quad
e^{2\gamma}=\frac{A^2-B^2+|C|^2}{16\s^2R_+R_-r_+r_-}, \nonumber\\
A&=&\s(R_++R_-)(r_++r_-)-\mu(R_+-R_-)(r_+-r_-), \nonumber\\
B&=&2m\s(r_++r_-), \quad C=2im\mu(R_+-R_-), \nonumber\\
I&=&2\s(R_++m)(R_-+m)-mz(R_+-R_-), \nonumber\\
R_\pm&=&\sqrt{\rho^2+(z\pm\s)^2},  \quad
r_\pm=\sqrt{\rho^2+(z\pm\mu)^2}, \quad \s=\sqrt{m^2+\mu^2}.
\label{mf2} \eea
By putting $\mu=0$ in (\ref{mf2}), one gets the Schwarzschild
solution.

\subsection{The pure vacuum limit}

In the absence of the magnetic dipole parameter $\mu$, the
solution (\ref{EF}) reduces to the $R=0$ special case of the
metric for two unequal counterrotating black holes \cite{CLM}. Our
alternative derivation of the solution \cite{CLM} permitted us to
work out a simple representation for the $R=0$ case which we give
below:
\bea f&=&\frac{A\bar A-B\bar B}{(A+B)(\bar A+\bar B)}, \quad
e^{2\gamma}=\frac{A\bar A-B\bar B}{{\cal K}_0\bar{\cal
K}_0R_+R_-r_+r_-}, \quad \omega=-\frac{2{\rm Im}[G(\bar
A+\bar B)]}{A\bar A-B\bar B}, \nonumber\\
A&=&(\s_1+\s_2)^2(R_+-R_-)(r_--r_+)
-4\s_1\s_2(R_+-r_-)(R_--r_+), \nonumber\\
B&=&2(m_1^2-m_2^2)[\s_2(R_--R_+)
+\s_1(r_+-r_-)], \nonumber\\
G&=&-zB +(m_1^2-m_2^2)[\s_1(R_++R_-)(r_+-r_-)
-\s_2(R_+-R_-)(r_++r_-) \nonumber\\ &&
-2\s_1\s_2(R_++R_--r_+-r_-)], \nonumber\\
R_\pm&=&\frac{m_1\mp\s_1-ia} {m_1\mp\s_1+ia}
\sqrt{\rho^2+(z\pm\s_1)^2}, \quad r_\pm=\frac{m_2\mp\s_2-ia}
{m_2\mp\s_2+ia} \sqrt{\rho^2+(z\pm\s_2)^2}, \nonumber\\
\s_1&=&\sqrt{m_1^2-a^2}, \quad \s_2=\sqrt{m_2^2-a^2}, \quad
K_0=4\s_1\s_2(m_1-m_2)^2/(m_1m_2), \label{mf3} \eea
and please note that the functions $R_\pm$ and $r_\pm$ are defined
here in a slightly different way than in (\ref{Rr}). Of course,
the above formulas (\ref{mf3}) are fully equivalent to the $\mu=0$
limit of the solution (\ref{EF})-(\ref{mf}).

We would like to remark that it was precisely the work \cite{CLM}
that motivated us to write this paper after we incidentally
discovered a misprint in the formula (38) of \cite{CLM} and then
took notice of a rather unusual property of that formula whose
correct form is
\be J_2=-\frac{J_1M_2}{M_1}\left(\frac{R+M_1-M_2}{R-M_1+M_2}
\right), \label{J2} \ee
where $M_i$ and $J_i$ are, respectively, the Komar \cite{Kom}
masses and angular momenta of the black-hole constituents, while
$R$ is the separation distance. Indeed, as it follows from
(\ref{J2}), for all $R>|M_1-M_2|$ the two Kerr black holes are
counterrotating; nevertheless, for $0\le R<|M_1-M_2|$ the black
holes become corotating, as the expression in parentheses on the
right-hand side of (\ref{J2}) then takes negative values.
Obviously, the authors of \cite{CLM} were only interested in the
configurations with $R>M_1+M_2$, when the counterrotating black
holes are separated by a massless strut, so that they discarded
other possibilities as unphysical or uninteresting. However, in
our opinion, it is the $R=0$ case that is probably most
interesting from  the physical point of view because this is the
only case of unequal constituents with equatorial symmetry, and
also because it might represent a legitimate final state of two
merging Kerr black holes. It is worth pointing out that the change
from counterrotation to corotation does not occur in the case of
equal black holes ($M_1=M_2$ in (\ref{J2})), so that the intrinsic
inequality of black holes is necessary for the formation of the
final configuration of corotating Kerr black holes described by
(\ref{mf3}). The change of the total angular momentum of the
system between its final ($R=0$) state and the initial state of
infinitely separated sources ($R=\infty$) is given by the simple
formula
\be \Delta J=2J_1M_2/M_1, \label{dJ} \ee
and we believe that this change of the total angular momentum,
which should certainly be attributed to the extremely strong
frame-dragging effects inside a larger black hole, could be
related to the production of relativistic jets in the centers of
galaxies.

\subsection{Magnetized Kerr solution}

By choosing $m_1=m$, $m_2=0$ in (\ref{EF}), we get a 3-parameter
variant of the magnetized Kerr spacetime of the form
\bea \E&=&\frac{A-B}{A+B}, \quad \Phi=\frac{C}{A+B}, \quad
A_\varphi={\rm Im}\left(\frac{I-zC}{A+B}\right), \nonumber\\
f&=&\frac{A\bar A-B\bar B+C\bar C}{(A+B)(\bar A+\bar B)}, \quad
e^{2\gamma}=\frac{A\bar A-B\bar B+C\bar C}{K_0\bar
K_0R_+R_-r_+r_-}, \quad \omega=-\frac{{\rm Im}[G(\bar A+\bar
B)+C\bar I]}{A\bar A-B\bar B+C\bar C}, \nonumber\\
A&=&\s_1\s_2(R_++R_-)(r_++r_-) -\s_2^2(R_+-R_-)(r_+-r_-) \nonumber\\
&&+ia[\s_1(R_++R_-)(r_+-r_-)-\s_2(R_+-R_-)(r_++r_-)],
\nonumber\\
B&=&2m\s_1[\s_2(r_++r_-) +ia(r_+-r_-)], \nonumber\\
C&=&2im\mu\s_2(R_+-R_-),
\nonumber\\
G&=&-2zB+\s_1\mu^2(R_++R_-)(r_+-r_-) -\s_2(2m^2+\mu^2)
(R_+-R_-)(r_++r_-) \nonumber\\
&& -2im^2a(R_+-R_-)(r_+-r_-)+2\s_2m\mu^2(R_+-R_-) \nonumber\\ &&
+2m\s_1(\mu^2-2a^2)(r_+-r_-) + 4ima\s_1\s_2(r_++r_-), \nonumber\\
I&=&-i\mu\{\s_1\s_2(R_++R_-)(r_++r_-) -\s_2^2(R_+-R_-)(r_+-r_-)
+ia[\s_1(R_++R_-)(r_+-r_-) \nonumber\\
&& -\s_2(R_+-R_-)(r_++r_-)] -2\s_1\s_2 [2(R_++m)(R_-+m)
-m(r_++r_-)] \nonumber\\
 &&-2ima[\s_2(R_+-R_-) -\s_1(r_+-r_-)]\}, \nonumber\\
R_\pm&=&\sqrt{\rho^2+(z\pm\s_1)^2},  \quad
r_\pm=\sqrt{\rho^2+(z\pm\s_2)^2}, \nonumber\\
\s_1&=&\sqrt{m^2-a^2+\mu^2}, \quad \s_2=\sqrt{\mu^2-a^2}, \quad
K_0=4\s_1\s_2. \label{mfMK} \eea

This solution is different from the known generalization of the
Kerr solution obtained by one of us more than two decades ago
\cite{Man2}. The difference is clearly seen if one considers the
axis data of the solution (\ref{mfMK}), namely,
\be e(z)=\frac{(z-m-ia)(z+ia)-\mu^2} {(z+m-ia)(z+ia)-\mu^2}, \quad
f(z)=\frac{im\mu} {(z+m-ia)(z+ia)-\mu^2}, \label{axisMK} \ee
where the magnetic dipole parameter $\mu$ enters the expressions
of both $e(z)$ and $f(z)$, unlike the solution \cite{Man2} whose
$e(z)$ coincides with the axis data of the Kerr metric \cite{Ker}
and hence is free of the magnetic parameter. However, the presence
of $\mu$ on the right-hand side of $e(z)$ in (\ref{axisMK}) is
quite acceptable as it is well known that magnetic field is able
to distort the stars \cite{BGo}, thus affecting the structure of
their gravitational multipoles.

\subsection{The extreme limit}

The extreme limit of the solution (\ref{EF}) corresponds to the
case of equal overlapping constituents, when $M_1=M_2$ and
$\s_1=\s_2$, and the application of the L'H\^opital rule to
formulas (\ref{EF}) and (\ref{mf}) is then required. By
introducing the spheroidal coordinates $x$ and $y$ via the
relations
\bea x&=&\frac{1}{2\s}(r_++ r_-), \quad y=\frac{1}{2\s}(r_+-r_-),
\quad r_\pm=\sqrt{\rho^2+(z\pm\s)^2}, \nonumber\\
\s&=&\sqrt{m^2-a^2+\mu^2}, \label{xy} \eea
it is possible to write down the resulting expressions in terms of
four polynomials $\lambda$, $\nu$, $\kappa$ and $\chi$. Thus, for
the potentials $\E$, $\Phi$ and $A_\varphi$ we get
\bea \E&=&\frac{A-B}{A+B}, \quad \Phi=\frac{C}{A+B}, \quad
A_\varphi={\rm Im}\left(\frac{I}{A+B}\right), \nonumber\\
A&=&\lambda^2+2m^2[\lambda +2ia\s xy(1-y^2)],
\nonumber\\
B&=&2m[(\s x+iay)\lambda+2im^2ay(1-y^2)], \nonumber\\
C&=&2i\mu my\lambda,
\nonumber\\
I&=&\frac{i}{2}\mu(1-y^2)[\kappa+4m^2(\s x+m-iay)^2]. \label{EFA3}
\eea
while the metrical fields $f$, $\gamma$ and $\omega$ are defined
by the expressions
\bea f&=&\frac{N}{D}, \quad e^{2\gamma}=\frac{N}{\s^8(x^2-y^2)^4},
\quad \omega=\frac{(y^2-1)W}{N}, \nonumber\\
N&=&\lambda^4-\s^2(x^2-1)(1-y^2)\nu^2, \nonumber\\
D&=&N+\lambda^2\kappa+(1-y^2)\nu\chi, \nonumber\\
W&=&\s^2(x^2-1)\nu\kappa+\lambda^2\chi, \nonumber\\
\lambda&=&\s^2(x^2-y^2)-m^2(1-y^2), \nonumber\\
\nu&=&4m^2ay^2, \nonumber\\
\kappa&=&4m[\s^2(\s x+2m)(x^2-y^2)+m^2\s
x(y^2+1)+m(2m^2+\mu^2)y^2], \nonumber\\
\chi&=&4ma[\s^2(\s x+2m)(x^2-y^2)+m^2\s x(1-y^2)]. \label{mfext}
\eea
Note that in the literature on exact solutions the polynomials
$\lambda$, $\nu$, $\kappa$ and $\chi$ have been previously used
exclusively in application to the metric functions $f$, $\gamma$
and $\omega$ \cite{Ern2,Per}, so that our paper actually pioneers
the use of these polynomials for getting a concise form of the
Ernst potentials $\E$ and $\Phi$ too. We also note that the
magnetic potential $A_\varphi$ can be written alternatively in the
form
\bea A_\varphi&=&\frac{2\mu(y^2-1)F}{D}, \nonumber\\
F&=&\lambda[\frac{1}{4}\kappa+m^2(\s x+m)^2-m^2a^2y^2]
[\lambda+2m(\s x+m)] \nonumber\\ &&-4m^3a^2y^2(\s x+m)
[\lambda+2m(\s x+m)(1-y^2)]. \label{A3} \eea

Remarkably, the vacuum ($\mu=0$) limit of the solution
(\ref{EFA3}) differs from the well-known Tomimatsu-Sato \cite{TSa}
and Kinnersley-Chitre \cite{KCh} solutions. This was a real
surprise to us since our initial intention was only to see how
this limit is contained in the Kinnersley-Chitre 5-parameter
family of solutions. In view of the potential interest the new
vacuum solution might represent, below we write it out explicitly:
\bea \xi&=&\frac{1-\E}{1+\E}= \frac{2m[(\s
x+iay)\lambda+2im^2ay(1-y^2)]} {\lambda^2+2m^2[\lambda +2ia\s
xy(1-y^2)]}, \nonumber\\ \lambda&=&m^2(x^2-1)-a^2(x^2-y^2), \quad
\s=\sqrt{m^2-a^2}. \eea
The corresponding metric functions are easily obtainable from
(\ref{mfext}).

\section{Concluding remarks}

We believe that the 4-parameter electrovacuum solution considered
in the present paper, as well as some of its particular limits,
provide interesting new opportunities for modeling the exterior
gravitational and electromagnetic fields of rotating bodies and
enlarge our knowledge about possible final states of two
interacting black holes. The solution has a clear physical
interpretation since it arises within a legitimate binary system
of counterrotating nonequal black holes, and the corotation of its
constituents, though unexpected at first glance but still natural,
should be attributed to strong dragging effects that involve a
smaller black hole in corotation with the larger one. We have
shown that the overlapping constituents in the case of two Kerr
black holes have a larger total angular momentum than at infinite
separation, the increase being defined by formula (\ref{dJ}), and
the latter formula also holds in the presence of the
electromagnetic field. It looks to us plausible to suppose that
the aforementioned change of the angular momentum could be related
to the mechanisms that are responsible for the production of
relativistic jets at the galactic nuclei \cite{BZn,Sha};
indirectly, this also suggests a recent article \cite{Rod} in
which the analysis of binary static systems of black holes with
magnetic charges has been considered as a right step towards a
better understanding of the jets phenomenon. We hope to be able to
shed more light on the details of that relation in the future.

As a final remark we would like to mention that the 4-parameter
solution considered in this paper can be trivially generalized to
include an additional parameter of electric dipole moment
$\varepsilon$ by the substitution $i\mu\to\varepsilon+i\mu$,
$\mu^2\to\varepsilon^2+\mu^2$, but we have not found for ourselves
any physical justification to do it here.

\section*{Acknowledgments}
This work was partially supported by CONACYT of Mexico, and by
Project FIS2015-65140-P (MINECO/FEDER) of Spain.

\newpage

\begin{figure}[htb]
\centerline{\epsfysize=100mm\epsffile{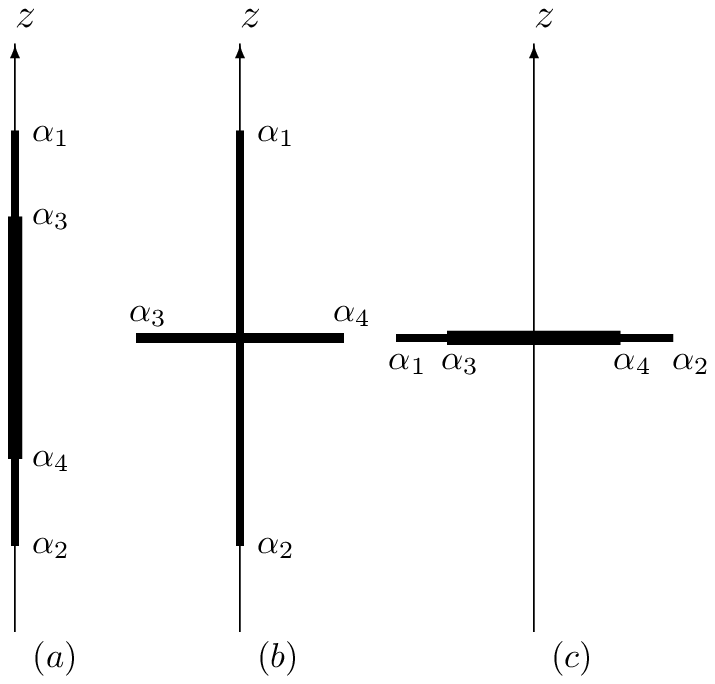}} \caption{Three
different types of systems with overlapping constituents: (a)
subextreme configuration, (b) subextreme-extreme configuration,
(c) hyperextreme configuration.}
\end{figure}

\end{document}